\documentclass[12pt]{article}

\usepackage{texdraw}
\usepackage[pdftex]{graphicx}
\normalbaselines
\begin{document}

%\hfill SEWT-4-ENG.tex % \psi_1  \leftrightarrow \psi_3

\begin{center}
{\Large \bf Higgsless Electroweak Theory following from  the Spherical Geometry }\\
%(31.05.2007)
\end{center}

\begin{center}
{\large  N.A. Gromov}\\
Department of Mathematics, \\
Komi Science Center UrD RAS \\
Kommunisticheskaya st., 24, Syktyvkar, 167982, Russia \\
E-mail: gromov@dm.komisc.ru
\end{center}

\begin{center}
{\bf Abstract}
\end{center}
A new formulation  of the Electroweak Model with  3-dimensional spherical geometry in the %matter fields 
 target space is suggested. The free Lagrangian in the spherical field space along with the standard gauge field Lagrangian form the full Higgsless  Lagrangian of the model, whose second order terms reproduce the same fields with the same masses as the Standard Electroweak Model. 
The vector bosons and electron masses are generated automatically, so there is no need in special mechanism.
  
\vspace{3mm}

%\vspace{5mm}
\section{Introduction }

The Standard Electroweak Model (SEWM) based on gauge group $ SU(2)\times U(1)$ gives a good  description of electroweak processes. One of the unsolved problems is the origin of electroweak symmetry breaking.  
In the standard formulation the scalar field (Higgs boson)  performs this task via Higgs mechanism,  which  generates a mass terms for vector bosons. However, it isnot yet  %has still not been 
experimentally verified whether electroweak symmetry is broken by such a Higgs mechanism, or by something else.

 The emergence  of  large number  Higgsless models  \cite{SCHKT-06}--\cite{CGPT-03} was stimulated by difficulties with Higgs boson. These models are mainly based on extra dimensions of different types or %more
 larger gauge groups. The  construction given in   \cite{S-07} is based on an observation: the underlying group of SEWM can be represented as a semidirect product of $U(1)$ and $SU(2).$

In the previous papers \cite{G-06-1},\cite{G-06-2}, where the gauge field theories based on  non-semisimple  contracted Cayley-Klein groups were considered, it was  noted that Higgs mechanism looks very artificial and Higgs boson being its  artefact  is unobservable. In the present paper a new formulation  of the Higgsless Electroweak Model with the 3-dimensional spherical geometry in the target space is suggested.

\section{Standard Electroweak Model }

The bosonic sector of SEWM %the Standard Electroweak Model 
is $SU(2)\times U(1)$ gauge theory in the space $\Phi_2({\bf C}) $  of fundamental representation of $SU(2).$ The  bosonic Lagrangian is given by the sum
\begin{equation}
L_B=L_A + L_{\phi},
\label{e1}
\end{equation}
where
\begin{equation}
L_A=\frac{1}{8g^2}\mbox{Tr}(F_{\mu\nu})^2-\frac{1}{4}(B_{\mu\nu})^2= -\frac{1}{4}[(F_{\mu\nu}^1)^2+(F_{\mu\nu}^2)^2+(F_{\mu\nu}^3)^2]-\frac{1}{4}(B_{\mu\nu})^2
\label{e8}
\end{equation}
is the gauge field Lagrangian for $SU(2)\times U(1)$ group and
\begin{equation}   
  L_{\phi}= \frac{1}{2}(D_\mu \phi)^{\dagger}D_\mu \phi -V(\phi)
\label{e9}
\end{equation}  
is the matter field Lagrangian. Here 
$ \phi= \left(
\begin{array}{c}
	\phi_1 \\
	\phi_2
\end{array} \right) \in \Phi_2({\bf C}),\;$
  $D_{\mu}$ are the covariant derivatives
 \begin{equation}
D_\mu\phi=\partial_\mu\phi -ig\left(\sum_{k=1}^{3}T_kA^k_\mu \right)\phi-ig'YB_\mu\phi,
\label{e11}
\end{equation} 
where $T_k=\frac{1}{2}\sigma_k,$ with $\sigma_k$ being Pauli matrices, are generators of $SU(2)$ and 
$Y=\frac{1}{2}{\bf 1}$ is generator of $U(1).$
The gauge fields 
\begin{equation}
A_\mu (x)=-ig\sum_{k=1}^{3}T_kA^k_\mu (x),\quad B_\mu (x)=-igYB_\mu (x)
\label{e2}
\end{equation} 
 take their values in Lie algebras $su(2),$  $u(1)$ respectively,  and the stress tensors are
\begin{equation}
F_{\mu\nu}(x)={\cal F}_{\mu\nu}(x)+[A_\mu(x),A_\nu(x)],\quad  B_{\mu\nu}=\partial_{\mu}B_{\nu}-\partial_{\nu}B_{\mu}.         
\label{e6}
\end{equation} 
The potential $V(\phi)$ in (\ref{e9}) is introduced by hand in a special form   %of mexican hat
\begin{equation}   
  V(\phi)=\frac{\lambda }{4}\left(\phi^{\dagger}\phi- v^2\right)^2, 
\label{e10}
\end{equation}  
where $\lambda, v $ are constants. 

The Lagrangian   $L_B$ (\ref{e1}) describe  massless fields. To generate   mass terms for the vector bosons without breaking the gauge invariance one uses the Higgs mechanism. One of  $L_B$ %(\ref{e1}) 
ground states
\begin{equation}   
  \phi^{vac}=\left(\begin{array}{c}
	0  \\
	v 
\end{array} \right), \quad  A_\mu^k=B_\mu=0
\label{e13}
\end{equation}
is taken as a vacuum state of the model, and small field excitations 
\begin{equation}   
 \phi_1(x), \quad \phi_2(x)=v+\chi(x), \quad A_\mu^a(x), \quad B_\mu(x)
\label{e14}
\end{equation}    
with respect to the vacuum  are regarded.
  %$$
 The matrix  
  $
  Q=Y+T^3=\left(\begin{array}{cc}
	1 & 0 \\
	0 & 0
\end{array} \right),
  $
  which annihilates the ground state $Q\phi^{vac}=0,$ is the generator of the electromagnetic subgroup
    $U(1)_{em}. $ 
The new fields 
  $$
  {W_\mu^{\pm}=\frac{1}{\sqrt{2}}\left(A_\mu^1\mp iA_\mu^2  \right)}, % \in {\bf C},
  $$
  %$$
\begin{equation}   
 { Z_\mu =\frac{1}{\sqrt{g^2+g'^2}}\left( gA_\mu^3-g'B_\mu \right)},\quad
 { A_\mu =\frac{1}{\sqrt{g^2+g'^2}}\left( g'A_\mu^3+gB_\mu \right)} % \in {\bf R}.
\label{e15}
\end{equation}    
  %$$  
are introduced, where  $W_\mu^{\pm} $ are complex  $(W_\mu^{-})^*=W_\mu^{+} $ and $Z_\mu, A_\mu $ are real. 

   The second order terms of the Lagrangian (\ref{e1}) are as follows 
  $$
  L_B^{(2)}=-{\frac{1}{2}{\cal W}_{\mu\nu}^{+}{\cal W}_{\mu\nu}^{-}+m_W^2W_\mu^{+}W_\mu^{-} } -{\frac{1}{4}{\cal F}_{\mu\nu}{\cal F}_{\mu\nu}}-
  $$
\begin{equation}   
   -{\frac{1}{4}{\cal Z}_{\mu\nu}{\cal Z}_{\mu\nu}+\frac{1}{2}m_Z^2Z_\mu Z_\mu} +  
  {\frac{1}{2}\left(\partial_\mu\chi \right)^2 -\frac{1}{2}m_{\chi}^2\chi^2},
\label{e16}
\end{equation} 
where 
$
{\cal Z}_{\mu\nu}=\partial_\mu Z_\nu-\partial_\nu Z_\mu, \; 
{\cal F}_{\mu\nu}=\partial_\mu A_\nu-\partial_\nu A_\mu, \; 
{\cal W^{\pm}}_{\mu\nu}=\partial_\mu W^{\pm}_\nu-\partial_\nu W^{\pm}_\mu. \; 
$   
This describes massive vector fields  {$W_\mu^{\pm}$ with identical mass  $m_W=\frac{1}{2}gv$} ($W$-bosons), massless  vector field {$A_\mu, \; m_{A}=0$} (photon),
massive vector field    {$Z_\mu$ with the mass  $m_Z=\frac{v}{2}\sqrt{g^2+g'^2}$} ($Z$-boson),
and massive scalar field {$ \chi,\; m_{\chi}=\sqrt{2\lambda}v$} (Higgs boson).

  $W$- and $Z$-bosons have been    observed and have the masses 
  $  m_W=80 GeV,$ $ m_Z=91 GeV.  $
 % However all attempts to verify experimentally  the scalar Higgs boson up to now failed.
 Higgs boson has not been  experimentally verified up to now.

\section{Higgsless Electroweak Model with 3D spherical matter  space}

The complex 2D space $\Phi_2$ can be regarded as 4D real Euclidean space ${\bf R}_4.$ Let us introduce the real fields $r, \bar{\psi}=(\psi_1,\psi_2,\psi_3)$ by  %as follows
\begin{equation}   
\phi_1=r(\psi_2+i\psi_1), \quad \phi_2=r(1+i\psi_3) .   
\label{e17}
\end{equation} 
It is easy to see that the quadratic form $\phi^\dagger\phi=\phi_1^*\phi_1+\phi_2^*\phi_2=R^2$ %in $\Phi_2$
 is invariant with respect to gauge transformations. For the real fields this form is written as
$r^2(1 + \bar{\psi}^2)=R^2, $ where $\bar{ \psi}^2=\psi_1^2+ \psi_2^2+\psi_3^2,$ therefore 
%From this it follows that
\begin{equation}   
r=\frac{R}{\sqrt{1 + \bar{ \psi}^2}}   
\label{e18}
\end{equation} 
Hence there are    %and we have
three independent real fields $\bar{\psi}.$ These fields  belong to the space $ \Psi_3 $
with noneuclidean spherical geometry which is realized  on the 3D sphere in 4D Euclidean  space ${\bf R}_4.$ The fields $\bar{\psi}$ are intrinsic Beltrami coordinates on $ \Psi_3. $

  The potential (\ref{e10}) is the constant  $V(\phi)=\lambda \left(R^2- v^2\right)^2/4  $ and $V(\phi)=0$ for   $R=v. $ Therefore, let us define  the {\it free} gauge invariant matter field Lagrangian $L_\psi $ with the help of the metric tensor
\begin{equation}
g_{kk}(\bar{\psi})=\frac{1+\bar{\psi}^2-\psi_k^2}{(1+\bar{\psi}^2)^2}, \quad
g_{kl}(\bar{\psi})=\frac{-\psi_k \psi_l}{(1+\bar{\psi}^2)^2} 
\label{p5}
\end{equation}     
of  %the spherical space 
$ \Psi_3 $ in the form   
\begin{equation}
L_\psi=\frac{R^2}{2}\sum_{k,l=1}^3g_{kl}D_\mu\psi_kD_\mu\psi_l=
\frac{R^2\left[(1+\bar{\psi}^2)(D_{\mu}\bar{\psi})^2-(\bar{\psi},D_{\mu}\bar{\psi})^2\right] }{2(1+\bar{\psi}^2)^2}.
\label{d7}
\end{equation}
The covariant derivatives  (\ref{e11}) are obtained using the representations of generators for the algebras $su(2),$ $u(1)$ in the space $\Psi_3$ \cite{Gr-07}
$$
T_1\bar{\psi}=\frac{i}{2}\left(\begin{array}{c}
-(1+\psi_1^2)	 \\
 \psi_3-\psi_1\psi_2\\
 -(\psi_2+\psi_1\psi_3)
\end{array} \right), \quad
T_2\bar{\psi}=\frac{i}{2}\left(\begin{array}{c}
-(\psi_3+\psi_1\psi_2) \\
 -(1+\psi_2^2)\\
 	\psi_1-\psi_2\psi_3
\end{array} \right),
$$
\begin{equation}
T_3\bar{\psi} =\frac{i}{2}\left(\begin{array}{c}
-\psi_2+\psi_1\psi_3	\\
 \psi_1+\psi_2\psi_3 \\
 1+\psi_3^2
\end{array} \right), \quad
Y\bar{\psi} =\frac{i}{2}\left(\begin{array}{c}
-(\psi_2+\psi_1\psi_3)	\\
 \psi_1-\psi_2\psi_3 \\
 -(1+\psi_3^2)
\end{array} \right),
 \label{p22}
\end{equation} 
and are as follows: 
$$
D_\mu \psi_1=\partial_\mu \psi_1+\frac{g}{2}\left[-(1+\psi_1^2)A_\mu^1 -(\psi_3+\psi_1\psi_2)A_\mu^2-(\psi_2-\psi_1\psi_3)A_\mu^3 \right] - \frac{g'}{2}(\psi_2+\psi_1\psi_3)B_\mu,
$$
$$
D_\mu \psi_2=\partial_\mu \psi_2+\frac{g}{2}\left[(\psi_3-\psi_1\psi_2)A_\mu^1 -(1+\psi_2^2)A_\mu^2 + (\psi_1+\psi_2\psi_3)A_\mu^3 \right] + \frac{g'}{2}(\psi_1-\psi_2\psi_3)B_\mu,
$$
\begin{equation}
D_\mu \psi_3=\partial_\mu \psi_3+\frac{g}{2}\left[-(\psi_2+\psi_1\psi_3)A_\mu^1 +(\psi_1-\psi_2\psi_3)A_\mu^2+(1+\psi_3^2)A_\mu^3 \right] - \frac{g'}{2}(1+\psi_3^2)B_\mu.
\label{d9}
\end{equation}  
%
%The gauge fields Lagrangian (\ref{e8}) is not changed since matter fields $\phi$ are not included in it.
%The gauge fields Lagrangian (\ref{e8}) is not changed since do not included  fields $\phi.$  
The gauge fields Lagrangian  does not depend on the fields $\phi$ and therefore remains unchanged (\ref{e8}).

For  small fields, the second order  Lagrangian (\ref{d7}) is written as
\begin{equation}  
L_\psi^{(2)}=\frac{R^2}{2}\left[(D_\mu \bar{\psi})^{(1)}\right]^2 = 
\frac{R^2}{2}\sum_{k=1}^3\left[(D_\mu \psi_k)^{(1)}\right]^2 , 
\label{e22-1}
\end{equation} 
where linear terms in covariant derivates (\ref{d9}) have the form %are 
$$
(D_\mu \psi_1)^{(1)}= \partial_\mu\psi_1-\frac{g}{2}A_\mu^1= -\frac{g}{2}\left(A_\mu^1-\frac{2}{g}\partial_\mu\psi_1\right) =-\frac{g}{2}\hat{A}_\mu^1,
$$
$$  
(D_\mu \psi_2)^{(1)}= \partial_\mu\psi_2-\frac{g}{2}A_\mu^2= 
-\frac{g}{2}\left(A_\mu^2-\frac{2}{g}\partial_\mu\psi_2\right)= -\frac{g}{2}\hat{A}_\mu^2, %\quad
$$
\begin{equation}
(D_\mu \psi_3)^{(1)}=\partial_\mu\psi_3+\frac{g}{2}A_\mu^3-\frac{g'}{2}B_\mu=
\partial_\mu\psi_3+\frac{1}{2}(gA_\mu^3-g'B_\mu)=\frac{1}{2}\sqrt{g^2+g'^2}Z_\mu.
\label{e22-2}
\end{equation} 
For the new fields 
$$ 
%\begin{equation} 
W^{\pm}_\mu = \frac{1}{\sqrt{2}}\left(\hat{A}^1_\mu \mp i \hat{A}^2_\mu \right), \quad (W^{-}_\mu)^*=W^{+}_\mu
% \mp \frac{\sqrt{2}}{g} 
%\left(\partial_\mu \psi_2 -i\partial_\mu \psi_3 \right),       
%\label{e24-5}
%\end{equation} 
$$
\begin{equation}  
 Z_\mu =\frac{gA^3_\mu-g'B_\mu + 2\partial_\mu \psi_3}{\sqrt{g^2+g'^2}}, \quad 
 A_\mu =\frac{g'A^3_\mu+gB_\mu}{\sqrt{g^2+g'^2}}
\label{e24}
\end{equation} 
Lagrangian (\ref{e22-1}) is rewritten as follows
\begin{equation}   L_{\psi}^{(2)}= %-\frac{1}{4}\sum_{k=1}^3 ({\cal F}^k_{\mu\nu})^2 - \frac{1}{4}(B_{\mu\nu})^2
  +\frac{R^2g^2}{4}W^{+}_\mu W^{-}_\mu +\frac{R^2(g^2+g'^2)}{8}\left(Z_\mu \right)^2.
\label{e25}
\end{equation} 

The quadratic  part of the full Lagrangian
\begin{equation}  
 L_B^{(2)}= -\frac{1}{2}{\cal W}^{+}_{\mu\nu}{\cal W}^{-}_{\mu\nu} + m_{W}^2W^{+}_\mu W^{-}_\mu
 - \frac{1}{4}({\cal F}_{\mu\nu})^2 -\frac{1}{4}({\cal Z}_{\mu\nu})^2 +\frac{m_Z^2}{2}\left(Z_\mu \right)^2,  
\label{e26}
\end{equation} 
where 
\begin{equation} 
m_W=\frac{Rg}{2}, \quad m_Z=\frac{R}{2}\sqrt{g^2+g'^2},
\label{e27}
\end{equation}
and  
$
{\cal Z}_{\mu\nu}=\partial_\mu Z_\nu-\partial_\nu Z_\mu, \; 
{\cal F}_{\mu\nu}=\partial_\mu A_\nu-\partial_\nu A_\mu, \; 
{\cal W^{\pm}}_{\mu\nu}=\partial_\mu W^{\pm}_\nu-\partial_\nu W^{\pm}_\mu, \; 
$    
 describes all the experimentally verified parts of SEWM %the Standard Electroweak Model 
 but does not include the scalar Higgs field. For $R=v$ the masses (\ref{e27}) are identical to those of the SEWM %Standard Electroweak Model 
  (\ref{e16}).

The fermion Lagrangian of SEWM %the Standard Electroweak Model 
is taken in the form \cite{R-99}
\begin{equation} 
L_F=L_l^{\dagger}i\tilde{\sigma}_{\mu}D_{\mu}L_l + e_r^{\dagger}i\sigma_{\mu}D_{\mu}e_r -
h_e[e_r^{\dagger}(\phi^{\dagger}L_l) +(L_l^{\dagger}\phi)e_r],
\label{e28}
\end{equation}
where
$
L_l= \left(
\begin{array}{c}
	\nu_{e,l}\\
	e_l^{-}
\end{array} \right)
$
is the $SU(2)$-doublet,  $e_l^{-} $ the $SU(2)$-singlet, $h_e$ is constant and $e_r, e_l, \nu_e $ are two component Lorentzian spinors. 
Here $\sigma_{\mu}$ are Pauli matricies, 
$\sigma_{0}=\tilde{\sigma_0}={\bf 1},$ $\tilde{\sigma_k}=-\sigma_k. $ 
The covariant derivatives $D_{\mu}L_l $ are given by (\ref{e11}) with $L_l$ instead of 
$\phi$ and $D_{\mu}e_r=(\partial_\mu + ig'B_\mu)e_r. $
 The convolution on the inner indices of $SU(2)$-doublet is denoted by $(\phi^{\dagger}L_l)$.

The matter fields $\phi$  appear in Lagrangian (\ref{e28}) only in mass terms. With the use of  (\ref{e17}) and (\ref{e18}), these mass terms are rewritten in the form
$$
h_e[e_r^{\dagger}(\phi^{\dagger}L_l) +(L_l^{\dagger}\phi)e_r]=
\frac{h_eR}{\sqrt{1+\bar{\psi}^2}}\left[e_r^{\dagger}e_l^{-} + e_l^{- \dagger}e_r + \right.
$$
\begin{equation} 
\left.  +i\psi_3\left(e_l^{- \dagger}e_r - e_r^{\dagger}e_l^{-} \right) + 
i\psi_1\left(\nu_{e,l}^{\dagger}e_r - e_r^{\dagger}\nu_{e,l} \right)+
i\psi_2\left(\nu_{e,l}^{\dagger}e_r + e_r^{\dagger}\nu_{e,l} \right)
\right].
\label{e29}
\end{equation}
Its  second order terms 
$ %\begin{equation} 
h_eR\left(e_r^{\dagger}e_l^{-} + e_l^{- \dagger}e_r \right)
%\label{e30}
$ %\end{equation}
provide  the electron mass $m_e=h_eR, $ and neutrino  remain massless.

\section{Conclusion}

The suggested  formulation  of the Electroweak Model with the gauge group $SU(2)\times U(1)$ based on
the 3-dimensional spherical geometry in the target space  describes all experimentally observed fields, and does not include the (up to now unobserved) scalar Higgs field. 
The {\it free} Lagrangian in the spherical matter field space is used instead of Lagrangian (\ref{e9}) with
the   %by hand introduced 
potential (\ref{e10}) of the special form (sombrero).
The  gauge field Lagrangian is the standard one. 
There is no need in Higgs   %special mass genereted 
mechanism since the vector field  masses are generated automatically.

The motion group of the spherical space $\Psi_3$ is isomorphic to $SO(4).$ It is  known \cite{G-07}, that
the bosonic sector of SEWM %the Standard Model 
has the custodial $SU(2)$-symmetry, and potential (\ref{e10}) is globally invariant with respect of $SO(4)$ which is  locally isomorphic to  the group 
$SU(2)_L\times SU(2)_R,$ where $SU(2)_L$ is the gauge group of SEWM %the Standard Model. 
The appearance of the extra $ SU(2)_R$-symmetry is interpreted as transformation of $SU(2)$-doublet by pseudoreal representation, that is the complex conjugate doublet is equivalent to the initial one. The introduction of the real fields $\bar{\psi}$ is  connected in a certain sense with $ SO(4)$-symmetry.

\section{Acknowledgments}

The author is    grateful to A.V. Zhubr and V.V. Kuratov for helpful discussions.

\end{document}